\documentclass[preprint,aps,showpacs,superscriptaddress,showkeys,amsmath]{revtex4}

\usepackage{graphicx}
\usepackage{dcolumn}
\usepackage{bm}
\usepackage{float}

\begin{document}


\title{The influence of intergranular interaction on the magnetization of the ensemble of oriented Stoner-Wohlfarth nanoparticles}


\author{A.A. Timopheev}
\email[e-mail: ]{ timopheev@iop.kiev.ua}
\affiliation{Institute of Physics NAS of Ukraine, Prospect Nauki str. 46, Kiev, 03028, Ukraine}

\author{S.M. Ryabchenko}
\affiliation{Institute of Physics NAS of Ukraine, Prospect Nauki str. 46, Kiev, 03028, Ukraine}

\author{V.M. Kalita}
\affiliation{Institute of Physics NAS of Ukraine, Prospect Nauki str. 46, Kiev, 03028, Ukraine}

\author{A.F. Lozenko}
\affiliation{Institute of Physics NAS of Ukraine, Prospect Nauki str. 46, Kiev, 03028, Ukraine}

\author{P.A. Trotsenko}
\affiliation{Institute of Physics NAS of Ukraine, Prospect Nauki str. 46, Kiev, 03028, Ukraine}

\author{V.A. Stephanovich}
\affiliation{Institute of Mathematics and Informatics, Opole University, Oleska 48, 45-052 Opole, Poland}

\author{A.M. Grishin}
\affiliation{Royal Institute of Technology, Electrum 229, S-164 40 Kista, Stockholm, Sweden}

\author{M. Munakata}
\affiliation{Energy Electronics Laboratory, Sojo University, Kumamoto 860-0082, Japan}


\date{\today}

\begin{abstract}

We consider the influence of interparticle interaction on the magnetization reversal in the oriented Stoner-Wohlfarth nanoparticles ensemble. To do so, we solve a kinetic equation for the relaxation of the overall ensemble magnetization to its equilibrium value in some effective mean field. Latter field consists of external magnetic field and interaction mean field proportional to the instantaneous value of above magnetization.  We show that the interparticle interaction influences the temperature dependence of a coercive field. This influence manifests itself in the noticeable coercivity at $T>T_{b}$ ($T_{b}$ is so-called blocking temperature). The above interaction can also lead to a formation of the "superferromagnetic" state with correlated directions of particle magnetic moments at $T>T_{b}$. This state possesses coercivity if the overall magnetization has a component directed along the easy axis of each particle. We have shown that the coercive field in the "superferromagnetic" state does not depend on measuring time. This time influences both $T_{b}$ and the temperature dependence of coercive field at $T<T_{b}$. We corroborate our theoretical results by measurements on nanogranular films (CoFeB)$_{x}$-(SiO$_{2}$)$_{1-x}$ with concentration of ferromagnetic particles close, but below  percolation threshold.

\end{abstract}


\maketitle

\section{\label{sec:intro}Introduction}

The consideration of interparticle interaction in an ensemble of single-domain superparamagnetic particles is important both from theoretical \cite{r1,r2} and practical points of view \cite{r3,r4,r5,r6,r7}. The above interaction alters (as compared to the case of noninteracting particles) the magnetization curves, the coercive fields and the temperature dependence of an ensemble magnetic susceptibility. The intergranular interaction is always of dipole nature, although there are cases where an additional exchange interaction occurs also.

To be more specific, if metallic ferromagnetic (FM) granules are embedded in conducting host, they can have exchange interaction via common electronic system of a composite even if the granules concentration is lower than the percolation threshold. In the latter case, the interaction is of RKKY type being small and of alternating sign. If the host is dielectric and the granules concentration is lower than the percolation threshold the interaction will be of entirely dipole nature. On the other hand, close to and/or above this threshold, the interaction between contacting granules may have exchange contribution of both RKKY and usual ion-ion types.

A possibility for homogeneous long-range FM order to appear due to presence of dipole-dipole interaction has been discussed by many authors beginning from Ref. \cite{r25}. It has been shown in Ref. \cite{r25}, that in simple cubic lattice of magnetic dipoles the latter interaction generates long-range antiferromagnetic order rather then FM one. The same conclusion has also been drawn in Ref. \cite{r26}. The question about stability of FM long-range order in the cases when corresponding static mean field solution predicts the appearance of FM order has been considered in Ref. \cite{r24}. This analysis shows that FM order is unstable with respect to 3D perturbations, but is stable with respect to 2D ones. Despite of the above discussions, the appearance of long-range FM order in granular systems with dielectric matrix has been detected experimentally in many systems with granules concentration both below and above percolation threshold. The properties of such FM ordered states of nanogranular systems have not been adequately explored.

Regardless of the intergranular interaction nature, the question about joint influence of superparamagnetism and intergranular ordering on the magnetic properties of FM particles ensembles is still opened. It is naturally to expect that intergranular interaction generates a correlation of the particles magnetic moment directions. If this interaction is of FM type, then below certain temperature $T_{sf}$, it should generate long-range magnetic order. Otherwise, so-called superspinglass state can be realized.

Most frequently, each particle has a certain crystallographic anisotropy. If its shape is non-spherical, the anisotropy can be of magnetostatic nature, related to demagnetization factors tensor for that shape. The thermal fluctuations lead to reorientation of a particle magnetic moment between several equivalent easy magnetization directions, dictated by the above anisotropy. In that case the observable magnetic properties of the ensemble are different depending on the relation between the reorientation time and the period of ensemble observation. For noninteracting particles, the thermally activated reorientations of their magnetic moments are mutually independent. Two kinds of ensemble superparamagnetic states can be distinguished. Namely, there are equilibrium and nonequilibrium superparamagnetic states. The equilibrium state occurs when the magnetic moment of an average particle (i.e. the typical ensemble particle with some average parameters) ``covers'' all permissible easy magnetization directions during the time of observation. The nonequilibrium or ``blocked'' superparamagnetic state occurs in the opposite case, when the particles are ``blocked'', i.e. they cannot alter their magnetization orientations during the observation time. The threshold temperature between these two states is called blocking temperature, $T_{b}$. The magnetic switching in the blocked state ($T<T_{b}$) has a hysteretic character, while at $T>T_{b}$ it is almost unhysteretic.

Now we ``turn on'' the interparticle interaction. If it has FM character, then at $T<T_{sf}$ ($T_{sf}$ is determined by the interaction and can be regarded as FM phase transition temperature) the particles magnetic moments influence each other. Their reorientations cannot be independent. In such case, the particles ensemble may be considered as an effective ferromagnet with hysteretic magnetization reversal. The origin of hysteresis is similar to ordinary ferromagnets - either a pinning of the domain walls motion (if the above structure has domains) or simply loss of stability of homogeneously magnetized ground state (see, e.g. Ref. \cite{r8} for details).

It is naturally that $T_{b}$ can vary depending on particles size, anisotropy and observation time. In turn, $T_{sf}$ should strongly depend both on particles size and mean interparticle distance. Hence, the situation when either $T_{b}>T_{sf}$ or $T_{b}<T_{sf}$ can occur. The main problem here is possible influence of above FM ordering on temperature dependence of a coercive field in superparamagnetic state. Other important question is how this influence is modified depending on observation time.

 The above ordered state has been identified for $T_{b}<T_{sf}$ in Refs. \cite{r9,r10,r11}. A nonzero, weakly temperature dependent coercive field has been observed. The authors \cite{r9,r10,r11} call this state ``superferromagnetic'', defining it as that lying in the temperature range between equilibrium (unhysteretic) and blocked (hysteretic) superparamagnetic states. Neither temperature nor observation time dependencies of a coercive field have been analyzed theoretically in Refs. \cite{r9,r10,r11}. The measurements in Refs. \cite{r9,r10,r11} were carried out on the films much thicker than the granules average size and with high enough relative volume content of the ferromagnetic component. It was supposed in Ref. \cite{r11} that intergranular interaction in their films has exchange nature. At the same time, in Refs. \cite{r12,r13}, where the ordering in above nanoparticles ensemble has also been observed, the absence of coercivity in the temperature range between $T_{sf}$ and $T_{b}$ has been reported. The coercivity appeared only at $T<T_{b}$ in contradiction with the data of Refs. \cite{r9,r10,r11}. We note that in Refs. \cite{r12,r13} the materials under investigation were the multilayer films rather then above 3D systems. In these materials, the layers of granules with relatively small 2D filling were separated by the insulating layers some thicker than the average size of a granule. It has been supposed in Refs. \cite{r12,r13} that such samples emulate 2D ensembles of particles with dipole intergranular interaction.

Thus, now there is a lack of complete understanding of the ``superferromagnetic'' state nature and its properties.

In this paper, we analyse theoretically and experimentally the influence of the interaction on the coercivity of an ensemble of oriented uniaxial superparamagnetic particles. We consider the above interaction in a mean field approximation \cite{r14,r15,r16} without discussing the nature of such interaction. The magnetization switching process is considered in the framework of Neel discrete orientations model \cite{r17} by the solution of a kinetic equation for magnetization similar to Refs. \cite{r15,r18,r19}. Contrary to those papers, here we discuss a possibility of superferromagnetic state creation. Also, here we study the changing of the temperature dependences of coercive field due to intergranular interaction with taking into account the influence of the observation time on the measured physical quantities. We consider the magnetization switching at linear (in time) magnetic field scanning. It corresponds to traditional scheme of magnetostatic measurements.

We show that at $T_{sf}>T_{b}$ the interparticle interaction generates coercivity at $T<T_{sf}$ and changes the ordinary (for a superparamagnet) temperature dependence of a coercive field at $T<T_{b}$. At $T<T_{b}$ the interaction yields the coercivity growth in coincidence with earlier Monte Carlo results \cite{r15,r20}. We have also shown that at $T<T_{b}$ the interparticle interaction modifies the temperature dependence of a coercive field, obeying Neel-Brown law.  One more result is that in ``superferromagnetic'' state the coercive field is related to the collective magnetization reversal of the particles and does not depend on measuring time.

 To corroborate the above theoretical results experimentally, we perform magnetostatic measurements in the granular (CoFeB)$_{x}$-(SiO$_{2}$)$_{1-x}$ films  \cite{r21,r22}. In the samples under investigation, the FM nanoparticles were anisotropic with easy magnetization axes oriented along a certain direction in a film plane. We study the temperature dependences of coercive field at different observation times. Our experimental results are in good coincidence with the theoretical model presented below.

\section{\label{sec:model}The model}

Let us consider an ensemble of interacting SW-particles with their easy axes pointing along the direction of an external magnetic field. For this case, the energy density per particle in a mean field approximation is:

\begin{equation}\label{EQ1}
U=-K\cos ^{2} (\theta )-m_{p} \left(H+\lambda m\right)\cos (\theta ).
\end{equation}

Here $K$ is a uniaxial magnetic anisotropy constant, $m_{p}$ is a single particle saturation magnetization (it is the same for each particle), $\theta$ is the angle between a particle magnetization vector and external magnetic field $H$ direction, $\lambda$ is a mean field interaction parameter and $m$ is an average magnetization per each ensemble particle. The latter quantity equals to the overall ensemble magnetization divided by the relative volume occupied by ferromagnetic particles in a sample. Below we will use the dimensionless magnetization $M=m/m_{p}$. This value will be the same both for the ensemble and for each single particle.

Let us pay attention that the potential energy profile (\ref{EQ1}) has the form of double-well potential. According to the Neel model \cite{r17} we can describe this system in the temperature range typical for the magnetostatic measurements ($0<T<3T_{b}$) as a system with two possible orientations of particles magnetic moments. If the magnetic field is directed along easy magnetization axes of the particles, the double well potential can be substituted by its two lowest energy levels so that our system can be described as a two-level system. In this case, the transitions between levels corresponding to magnetic moment reorientations occur as thermally activated hops over energy barrier.

In this model, the magnetization dynamics is of purely relaxational type. This means that time dependence of magnetization $M$ at fixed temperature $T$ and magnetic field $H$ can be described (similar to Refs. \cite{r15,r19}) by Bloch-like equation for z-component of magnetization only. If the interaction term $\lambda m$ is absent in Eq. (\ref{EQ1}), the equation for magnetization dynamics has the form:

\begin{equation} \label{EQ2}
\frac{\partial M(t)}{\partial t}=\frac{1}{\tau }\left[M_{\infty }-M(t)\right],
\end{equation}

where ${M_{\infty}}\equiv M(t\rightarrow \infty)$ is the equilibrium magnetization at fixed $H$ and $T$ and $\tau$ is the relaxation time. For our case of two-level system $\tau ^{-1}=W_{12} +W_{21}$, where $W_{ij}(j=1,2)$ are probabilities of transition between $i$ and $j$ levels in the double-well potential (\ref{EQ1}). According to approach \cite{r17} for SW particles, the final form of $\tau$ reads:

\begin{equation} \label{EQ3}
\tau = \frac{1}{f_{0} \left(\exp(-\frac{E_{b}-E_{1}}{kT})+\exp(-\frac{E_{b}-E_{2}}{kT})\right)}.
\end{equation}

Here $k$ is Boltzmann constant, $f_{0}\sim10^{8}\div10^{12}$ s$^{-1}$ for typical magnetic particles and $E_{1}$, $E_{2}$, $E_{b}$ are, respectively, the energies of minima of $U(\theta)$ and a barrier between them. The quantities $E_{1}$, $E_{2}$ and $E_{b}$ depend on $K$, $\lambda$ and $M(t)$. They are the functions of time by virtue of $M(t)$ dependence. The expressions (\ref{EQ2}) and (\ref{EQ3}) correspond to the approach, where a fictitious particle (corresponding to magnetization) is localized exactly in the minimum of U(\textit{q}) rather then ``smeared'' by temperature in a wide interval of angles $\theta$ within the well of the potential (\ref{EQ1}). For magnetic field sweeping times, taking place in magnetostatic measurements, this approach is well satisfied in the temperature range $T<(4\div6)T_{b}$, i.e. in the entire temperature domain. Thus magnetization reversal occurs by thermoactivation overbarrier hopping. In a mean field approximation, the equilibrium magnetization for such two-level system is determined by usual equation:

\begin{equation} \label{EQ4}
M_{\infty } =\tanh \frac{m_{p}(H+\lambda m_{\infty })V_{p}}{kT} =\tanh\frac{2(h+\lambda _{red}M_{\infty})}{T_{red}}.
\end{equation}

Here we introduce following dimensionless parameters: the dimensionless magnetic field, $h=H/H_{a}$ ($H_{a}=2K/m_{p}$ is the anisotropy field), the dimensionless relaxation time $\tau_{r}=\tau f_{0}$, the temperature $T_{red}=kT/(K V_{p})$ ($V_{p}$ is SW-particle volume) and the dimensionless parameter of interparticle interaction $\lambda _{red} =\lambda   m_{p}^{2} /(2K)$. We also introduce the dimensionless energy minima $E_{1}/(K V_{p})$, $E_{2}/(K V_{p})$, barrier maximum energy $E_{b}/(K V_{p})$, dimensionless time $t_{r}=t   f_{0}$ and time of measurements $t_{reg} =t_{exp}   f_{0}$. Here $t$ is a real dimensional time and $t_{exp}$ is a characteristic dimensional time of measurements, i.e.``measuring time'' - the time required for magnetic field sweeping in the range of $H_{a}$.

The introduction of the interaction term $\lambda m$ in the Eq.(\ref{EQ1}) modifies the character of relaxation. Namely, under magnetization reversal this term becomes time dependent as it comprises the magnetization $m(t)$. This means that the overall magnetization relaxes not to the above real equilibrium magnetization $M_{\infty}=M_{\infty}(H,T)$, but to certain (so far unknown) self-consistent equilibrium magnetization value, dictated by the effective magnetic field $H+\lambda m(t)$ at each time point. We denote this new hypothetical equilibrium magnetization as $m_{\infty }^{*}(t)$, and its normalized value as $M_{\infty }^{*}(t_{r})=m_{\infty}^{*}(t_{r}/f_{0})/m_{p}$. Here we note, that $M_{\infty }^{*}(t_{r})$ = $M_{\infty }^{*}[H,T,M(t_{r})]$ so that the kinetic equation for magnetization assumes the form:

\begin{equation} \label{EQ2a}
\frac{\partial M(t_{r})}{\partial t_{r}}=\frac{1}{\tau_{r}}[M_{\infty }^{*}-M(t_{r})], \tag{2a}
\end{equation}

where $M_{\infty }^{*}(t_{r})$ is determined by the equation

\begin{equation} \label{EQ4a}
M_{\infty }^{*}(t_{r})=\tanh\frac{m_{p}[H+\lambda m(t_{r}/f_{0})]V_{p}}{kT} =\tanh\frac{2[h+\lambda _{red} M(t_{r})]}{T_{red}}.\tag{4a}
\end{equation}

The equation (\ref{EQ2a}) with respect to (\ref{EQ4a}) will be solved numerically. To model the hysteresis loops, we consider, similar to Ref. \cite{r19}, the linear field sweep $h(t_{r})=(t_{r}/t_{reg})-1$ as it realized in real experiments. In the dimensionless variables the equation (\ref{EQ2a}) assumes the form:

\begin{eqnarray} \label{EQ5}
\frac{\partial M(t_{r})}{\partial t_{r}}=\left(\exp\left[-\frac{(t_{r}/t_{reg}-2+\lambda _{red} M(t_{r}))^{2}}{T_{red}}\right]+\exp\left[-\frac{(t_{r}/t_{reg}+\lambda _{red} M(t_{r}))^{2}}{T_{red}}\right]\right )\nonumber\\
\times
\left(\tanh\left[{\frac{2(t_{r}/t_{reg}-1+\lambda _{red}\cdot M(t_{r}))}{T_{red}}}\right]-M(t_{r})\right).
\end{eqnarray}

The last brackets in the right-hand side of Eq. (\eqref{EQ5}) define the difference between self-consistent hypothetical equilibrium magnetization \eqref{EQ4a} and its current value.

\section{\label{sec:resmod}Results of modeling}

\begin{figure}
\begin{center}
\includegraphics[width=8.5cm]{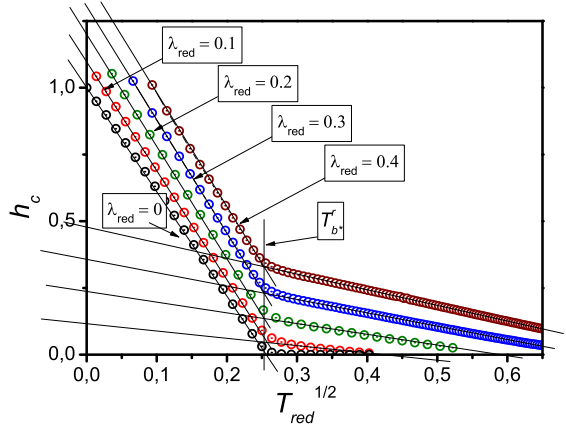}
\end{center}
\caption{\label{fig1} (color online) The temperature dependence of coercive field $h_{c}(\sqrt{T_{red}},t_{reg},\lambda _{red})$ for $t_{reg}=10^{8}$ and different values of $\lambda_{red}$ (0, 0.1, 0.2, 0.3, 0.4). Points are the results of numerical solution of Eq. (\ref{EQ5}) and the full lines correspond to low- and high temperature extrapolations of the numerical curves.}
\end{figure}

The solution of Eq. (\ref{EQ5}) shows that the account for the intergranular interaction term $\lambda _{red} M(t_{r})$ increases the coercivity. Also, the hysteresis loops become ``more rectangular'' ("harder"). The temperature dependences of the coercive field $h_{c}$, extracted from the hysteresis loops calculated with the help of Eq. (\ref{EQ5}), are shown on Fig. \ref{fig1} at different values of interaction parameter $\lambda _{red}$. It is seen that the interaction increases the coercive field. Besides that, at $2\lambda_{red}>T_{b}^{*}$ the dependence $h_{c}(T_{red})$ has two linear in $\sqrt{T_{red}}$ parts.

The low-temperature part is similar to Neel-Brown law:

\begin{equation} \label{EQ6}
 h_{c}(T_{red} )=1-\sqrt{T_{red}/T_{b}^{*}}  \hspace{15pt}\textrm{ for }   T_{red} <T_{b}^{*},
 \end{equation}

where $T_{b}^{*}$ is a dimensionless effective blocking temperature, determined by extrapolation of the low-temperature part of dependence $h_{c}(\sqrt{T_{red}})$ up to its intersection with abscissa axis. The value of $T_{b}^{*}$ depends on $\lambda _{red}$.

The high-temperature part of $h_{c}(\sqrt{T_{red}})$ lies at $T_{red} >T_{b}^{*}$. Its origin is a consequence of formation (due to interaction $\lambda _{red}>T_{b}^{*}/2$) of the state with correlated directions of granules magnetic moments realized at $T_{red}<T_{ord}=2\lambda_{red}$.  Here $T_{ord}$ corresponds to dimensionless temperature of the long range magnetic ordering in the granules ensemble. In Section \ref{sec:intro} (Introduction), the real (dimensional) temperature of such ordering had been denoted as $T_{sf}$.

Strictly speaking, at $\lambda _{red}<T_{b}^{*}/2$ and close to $T_{red}=2\lambda_{red}$, the dependence $h_{c}(\sqrt{T_{red}})$ acquires additional slope as compared to that in Eq. (\ref{EQ6}). This deviation, however, cannot be seen in the scale of Fig. \ref{fig1} so we do not plot corresponding curve on Fig. \ref{fig1}.

In the temperature range above blocking temperature, the relaxation time is much less than measuring time so that the appearance of coercivity in this temperature range is not a consequence of slow system response to magnetic field sweep. The coercivity at $T_{b}^{*}<T_{red}<T_{ord}$ is due to emergence of a self-consistent mean field of interparticle interaction at $T_{red}<T_{ord}$. The direction of latter field is the same as a direction of the external magnetic field at the initial magnetization stage, when we lower the magnetic field from saturation down to zero. This mean field stabilizes the directions of particles magnetic moments opposite to the external field direction during field scanning process in the interval $0<h<h_{c}$. In the above temperature range $T_{red}<T_{ord}$ the overall particles magnetization can be well described by Eq. \ref{EQ4}. This situation corresponds to the joint action of external and above self-consistent fields on each particle. In this temperature range, the quantity $|h_{c}|$  is determined by stability limits of Eq. \ref{EQ4} solution at opposite signs of $h$ and $\lambda_{red}M$. The values of $h$, where $\partial M/\partial h$ diverges, permit to obtain the expression for the coercive field $h_{c}^{int}$ related to interaction term. It reads:

\begin{equation} \label{EQ7}
h_{c}^{int}(T_{red},\lambda _{red})=\lambda _{red}\sqrt{1-T_{red}/T_{ord}}+\frac{T_{red}}{4}\log\left[\frac{1-\sqrt{1-T_{red}/T_{ord}}}{1+\sqrt{1-T_{red}/T_{ord}}}\right].
\end{equation}

\begin{figure}
\begin{center}
\includegraphics[width=8.5cm]{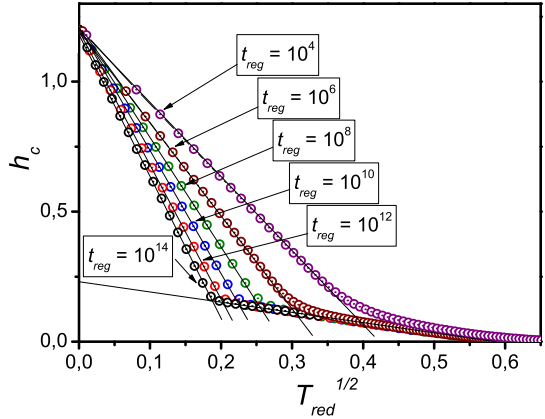}
\end{center}
\caption{\label{fig2} (color online) The temperature dependencies of coercive field $h_{c}(\sqrt{T_{red}},t_{reg},\lambda_{red})$ for $\lambda_{red}=0.2$ with different measuring times ($t_{reg} = 10^{4},10^{6},10^{8},10^{10},10^{12},10^{14}$ respectively). Points and full lines are the same as on Fig.\ref{fig1}.}
\end{figure}

Here $T_{ord}=2\lambda _{red}$ is the ordering temperature. In other words, this is a temperature where the state with correlated directions of particle magnetizations emerges. If $T_{ord} >T_{b}^{*}$, the above ordering occurs at $T_{b}^{*}<T<T_{ord}$, corresponding to the parameters values shown on Fig. \ref{fig1} and Fig. \ref{fig2}. The curve $h_{c}^{int}(T_{red})$ determined by Eq. (\ref{EQ7}) has quite complex shape, but it is proportional to $\sqrt{T_{red}}$ at temperatures from $0.1 T_{ord}$ to $0.9 T_{ord}$. In this temperature interval, the values of $h_{c}^{int}$ (Eq. (\ref{EQ7})) can be made equal to those from Fig. \ref{fig1} at $T_{red} >T_{b}^{*}$. Note that linear in $\sqrt{T_{red}}$ asymptotics of $h_{c}^{int}(T_{red})$ crosses the ordinate axis near $\lambda _{red}$.

More detailed analysis of calculated curves presented on Fig.\ref{fig1} shows that they can be described with good accuracy by the sum of Neel-Brown type contribution $h_{c}^{NB*}=1-\sqrt{T_{red}/T_{b*}^{r}}$ and the contribution $h_{c}^{int}$ determined by Eq. (\ref{EQ7}). Here $T_{b*}^{r} =T_{b*}^{r}(t_{reg})$ is a blocking temperature similar to that in Neel-Brown formula but modified with respect to real experimental conditions (continuous magnetic field sweeping for magnetization reversal curves registration, see \cite{r23} for details). The $T_{b*}^{r}$ value does not depend on $\lambda _{red}$.

The dependencies $h_{c}(\sqrt{T_{red}})$ for different measuring times $t_{reg}$ at fixed interaction parameter $\lambda _{red}$ are reported on Fig. \ref{fig2}. One can see that the low-temperature part of the curves depends on measuring time owing to the corresponding dependence of parameter $T_{b*}^{r}$. This is a characteristic feature of the systems without interaction and with thermally activated hopping at finite measuring time. The high-temperature side of the curves does not depend on $t_{reg}$. In this temperature region, all curves for different $t_{reg}$ have the same slope.

Note that for $T_{red}<T_{ord}$ the hysteresis curves have a "hard" (rectangular) shape. At the same time, for the case $\lambda _{red} = 0$ and $T_{red}$ smaller, but close to $T_{b*}^{r}$, the hysteresis loops have ``softer'' or ``more canted'' shape.

It is significant that the independence of a coercive field from measuring time at temperatures higher then blocking temperature is a consequence of quick (during measuring time or sooner) establishing of the equilibrium population in double-well potential of the SW-particles in this temperature range. More detailed analysis of our model shows that a faint dependence $h_{c}(t_{reg})$ is still present in this temperature range. Additionally, in this temperature range, the magnetization does not have a step at $h=h_{c}$ but rather varies continuously, changing its sign during the temperature dependent relaxation time $\tau_{r}$. At $T_{red} >T_{b*}^{r}$ the relaxation time $\tau_{r}$ is much shorter than measuring time. Note also, that our model with interparticle interaction does not imply the coercivity for magnetization along a difficult direction.
Thus, for ensemble of interacting SW-particles a temperature dependence of coercive field has the form:
\begin{equation} \label{EQ8}
h_{c}(T_{red},\lambda _{red})\cong h_{c}^{int}(T_{red},\lambda _{red})+(1-\sqrt{T_{red}/T_{b*}^{r}})
\end{equation}
for $T_{red}<T_{b*}^{r}$ and $h_{c}(T_{red},\lambda _{red})\cong h_{c}^{int}(T_{red},\lambda _{red})$ for $T_{red}>T_{b*}^{r}$, where $h_{c}^{int}(T_{red},\lambda _{red})$ determined by Eq. (\ref{EQ7}).  Note, that the Eq. (\ref{EQ8}) is exact everywhere, except for narrow region near $T_{b*}^{r}$ where it is fulfilled approximately.

\section{\label{sec:exp}Experimental}

To corroborate the above theoretical approach, we measure the magnetostatic characteristics of nano-granular films (Co$_{0.25}$Fe$_{0.66}$B$_{0.09}$)$_{x} - $(SiO$_{2}$)$_{1-x}$. The aim was to check the transition of SW particles ensemble from relaxation regime of magnetization reversal to steady-state regime of the intergranular ``ferromagnetic'' ordering (arising due to the interaction), when the coercive field ceases to depend on measuring time. In our measurements, we use the (Co$_{0.25}$Fe$_{0.66}$B$_{0.09}$)$_{x} - $(SiO$_{2}$)$_{1-x}$ films grown in the Energy Electronics Laboratory, Sojo University, Japan. The ferromagnetic granules were amorphous and their shape was close to the spherical. A strong easy-plane anisotropy related to the demagnetization factor arose for the entire film sample. In the granular films under investigation, the uniaxial anisotropy in a film plane had been formed by special technological measures \cite{r21,r22}. This anisotropy was supposedly related to the small deviation of the shape of granules from the spherical one. Thus the easiest (i.e. easy in a film plane) axes of all granules have been oriented almost parallel to each other. Therefore such granular system can be considered as an ensemble of easy axis oriented Stoner-Wohlfarth particles. The ensemble can be considered as ``noninteracting'' one for $x$ substantially lower than the percolation threshold $x_{c}$ and as ``strongly interacting'' one for $x\gg x_{c}$.
\begin{figure}
\begin{center}
\includegraphics[width=8.5cm]{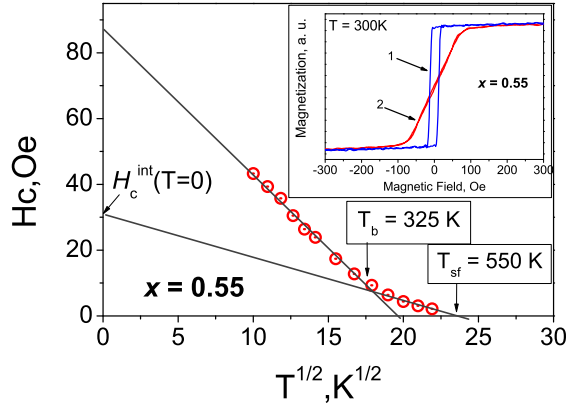}
\end{center}
\caption{\label{fig3} (color online) The temperature dependence of coercive field in the sample with $x=0.55$ measured along easy direction in a film plane. Inset reports the magnetization reversal curves along easy (1) and hard (2) directions in the same film plane at $T=300$ K.}
\end{figure}
According to the technologist, who had fabricated our sample, the ferromagnetic component content in it is $x=0.55$. It corresponds to volume fraction of the ferromagnetic granules about 0.26. It is strictly lower than volume fraction of percolation threshold. However, the film saturation magnetization, ferromagnetic resonance and granule magnetization \cite{r27} data have shown that real volume fraction of granules in this sample is essentially higher (then 0.26), up to $0.4\div 0.45$. We assert that this real volume fraction is a little below percolation threshold. To prove this assertion, we had measured the magnetoresistance curves at $T$=300K. The sample has high enough specific resistance, $\rho$($T$=300 K) = 250 mOhm/cm. The measured magnetoresistance curves contained the contribution from only tunneling magnetoresistance and did not contain the contribution from the anisotropic magnetoresistance. Since latter contribution in such films appears for $x>x_{c}$ only \cite{r22}, this result proves above assertion. For the above sample, the experimental data for magnetization in the film plane along easy direction (curve 1) at room temperature are presented on the inset to Fig. \ref{fig3}. One can see that the hysteresis loop is close to rectangular and has almost 100\% remanence. The field dependence of magnetization along a hard direction in a film plane (curve 2), has no hysteresis, has jogs at the intraplane anisotropy fields and is almost linear in magnetic field between jogs. The temperature dependence of coercivity (Fig. \ref{fig3}) demonstrates two linear in $\sqrt{T}$ parts, which coincide with the results of above theoretical modeling.

\begin{figure}
\begin{center}
\includegraphics[width=13cm]{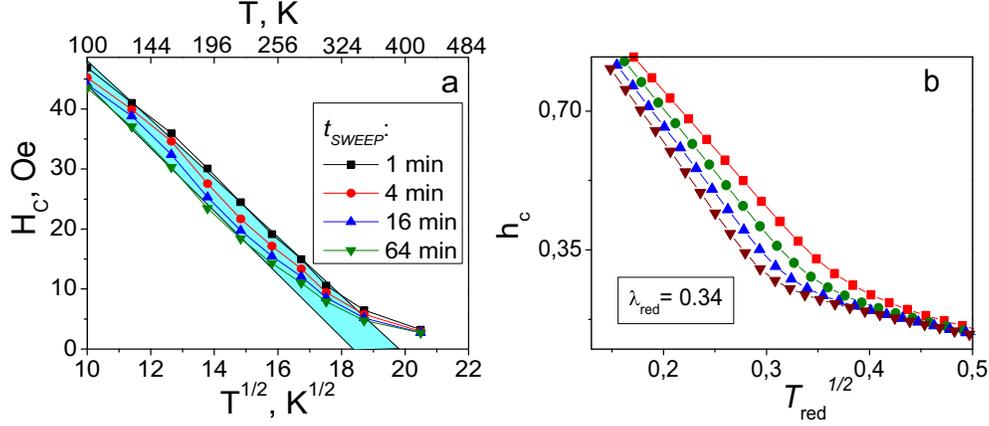}
\end{center}
\caption{\label{fig4} (color online) a) The temperature dependencies of coercive field $H_{c}$ for the sample with $x=0.55$ versus $\sqrt{T}$. The legend shows the field sweeping time from -500 Oe to +500 Oe and vice versa ($t_{sweep}$). b) The results of calculations of coercive field $h_{c}(\sqrt{T_{red}},t_{reg},\lambda _{red})$ for $\lambda _{red}=0.075$. The ratios of dimensionless measuring times $t_{reg}$ are the same as those on Fig. \ref{fig4}a.}
\end{figure}

At the same time, such dependence can be interpreted as a consequence of a bimodal size distribution of ensemble particles so that each particle group has its own $T_{b}$ value. To prove or disprove such possibility, we perform the number of magnetostatic measurements with different rates of magnetic field scanning. Fig. \ref{fig4}a reports the temperature dependences of coercive field $H_{c}$ at temperatures from 100 to 470 K at magnetization along easy direction. At temperatures lower than 100 K the temperature dependence of a coercive field in this film demonstrates an anomaly. We will not discuss that in the present paper, having restricted ourselves by temperature region above 100 K only.

The dependences $H_{c}(T)$ (in the form $H_{c}(\sqrt{T})$) on Fig.\ref{fig4}a are obtained for different times of a magnetic field sweeping ($t_{sweep}$=1 min, 4 min, 16 min, 64 min) from -500 Oe to +500 Oe and vice versa.  It is seen, that increasing of measuring time ($t_{exp}$ is proportional to $t_{sweep}$; actually for the anisotropy field value of this sample $H_{a}=80$ Oe, $t_{exp}=t_{sweep}/25$) leads, as it should be, to decreasing of $H_{c}$ at fixed temperature. It also yields the lowering of the blocking temperature $T_{b}$ determined as an intersection point of an asymptote to the linear low-temperature part of dependence $H_{c}(\sqrt{T})$ and abscissa axis. It follows from Fig. \ref{fig4}a that the dependences $H_{c}(\sqrt{T})$ behave like theoretical dependences from Fig. \ref{fig2}. In both experimental and simulation curves we observe a noticeable coercivity practically independent of measuring time above blocking temperature. The main properties of the presented curves can be formulated as follows. At low temperatures, $T\ll T_{b}$ ($T_{b}$ is taken for the \textit{longest} possible measuring time) the dependence of $H_{c}$ on measuring time becomes stronger with temperature increase. At a certain temperature, slightly lower then $T_{b}(t_{exp})$, a sensitivity of $H_{c}$ to measuring time variations reaches a maximum. At last, at $T>T_{b}$ (now $T_{b}$  is taken for the \textit{shortest} possible measuring time) the $H_{c}$ value ceases to depend on measuring time. At all temperatures a hysteresis loops remain "hard", conserving a rectangular form.

To compare the experimental and theoretical dependences of $H_{c}$ on measuring time it is necessary to account for the fact that measuring times, corresponding to the curves on Fig. \ref{fig2}, have 10 orders of magnitude variation, while experimental data from Fig. \ref{fig4}a have only 64 times difference. To illustrate the similarity between experimental and theoretical data, on Fig. \ref{fig4}b we present a number of theoretical curves with the relation of measurement times, identical to that in experiment. The curves are plotted for $\lambda_{red}=0.344$, corresponding to $\lambda=0.017$, CoFeB granule magnetization $m_{p}=1590$ Gauss \cite{r27} and in-plane ahisotropy field $H_{a}=2K/m_{p} = 80$ Oe. Latter value follows from the magnetization curve of our film in the ``hard--in-plane'' direction (curve 2 on inset to Fig. \ref{fig3}). The value $\lambda=0.017$ is obtained from the equation $H_{c}^{int}(T\to 0)=\lambda m_{p}$ (with respect to the value $H_{c}^{int}(T\to 0)= 30$ Oe following from Fig. \ref{fig3}), where $H_{c}^{int}$ is dimensional value of $h_{c}^{int}$. The values of $t_{reg}$ (shown on the legend to Fig. \ref{fig4}b) have been chosen from the condition of best fit between model and experimental $T_{b}(t_{reg})$ values.

\begin{figure}
\begin{center}
\includegraphics[width=13cm]{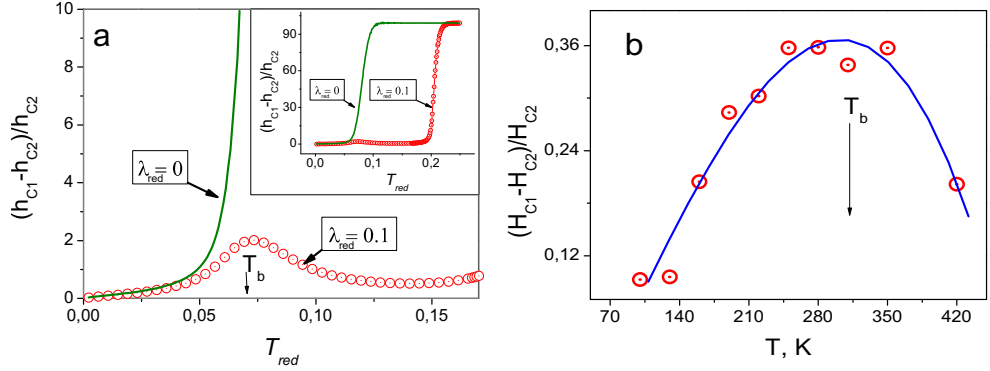}
\end{center}
\caption{\label{fig5} (color online) The temperature dependence of relative coercive field increments at registration time variations: a) - theoretical results; $h_{c1}(T_{red})$ is taken for $t_{reg}=10^{6}$ and $h_{c2}(T_{red})$ for $t_{reg}=10^{8}$. Full line corresponds to noninteracting case, circles - to interacting. Inset shows the same for wider temperature range. b) - the symbols represent experimental points, the curve is guide for eye; $H_{c1}(T)$ is taken for $t_{sweep}= 1$ min and $H_{c2}(T)$ for $t_{sweep}= 64$ min.}
\end{figure}

Although the reported experimental results are not quite identical to the results of our calculation, there is obvious qualitative coincidence. Namely, in our opinion, they demonstrate uniquely the existence of superferromagnetic state with coercivity. This fact is also corroborated by Fig. \ref{fig5}, where the temperature dependence of relative coercivity increments is reported for different measuring times. We define the above increments as the difference of coercive fields for two substantially different measuring times divided by the coercive field value at larger time. This dependence is reported both for theoretical (Fig.\ref{fig5}a) and experimental (Fig. \ref{fig5}b) results. It is seen, that the theoretical dependence for noninteracting ensemble has a sharp increase near the blocking temperature. At the same time, for the interacting ensemble, this dependence has a peak near $T_{b}$ with subsequent decrease. The experimental dependences (see Fig. \ref{fig5}b) also demonstrate the maximum with decrease.

Thus it turns out that the behaviour of experimentally observed $H_{c}(T, t_{sweep})$ dependence is qualitatively similar to results of our modeling for the case when intergranular interaction generates the state with correlated directions of particles magnetic moments and with a coercive field independent from measuring time. The above results illustrate the case when temperature $T_{sf}$ (``dimensional $T_{ord}$'') of transition to such "superferromagnetic" state exceeds blocking temperature $T_{b}$. The decreasing of intergranular interaction parameter $\lambda$ can lead to opposite situation, when $T_{sf}<T_{b}$. In this case the variations of dependence $H_{c}(\sqrt{T})$ also occur near $T_{sf}$. However, they are too faint to be observed experimentally.

It is possible to predict, that relaxational magnetization of the SW particles ensemble weakens if interaction energy exceeds the anisotropy energy. In this case the ensemble behaves as a uniform ferromagnetic medium with possible occurrence of "superdomains" consisting of many adjacent particles.

Thus, the experimental dependence $H_{c}(T)$ in nanogranular magnetic film with granules concentration close but a bit lower than percolation threshold is in qualitative agreement with model predictions for the system of superparamagnetic particles with intergranular interaction described in a mean field approximation. Additionally, the similar results have been obtained in our studies of the (Co$_{0.25}$Fe$_{0.66}$B$_{0.09}$)$_{x} - $(SiO$_{2}$)$_{1-x}$ film with nominal value $x=0.60$.

Note that in spite of aforementioned qualitative resemblance of the experimental and model data the quantitative correspondence is not so good. The obtained $t_{reg}$ values on Fig. \ref{fig4}b are too small.  It is the consequence of big granule size, which is needed for coincidence of the observed $T_{sf}$ and $T_{b}$ values with those expected from the model. Particularly, to coincide the $V_{p}$ value from the expression $k T_{sf} =\lambda m_{p}^{2} V_{p}$ with $\lambda$ obtained from the condition $H_{c}^{int}(T\to 0)=\lambda m_{p}\approx 30$ Oe, we need the mean diameter of granule 14 nm. To coincide the observed values of $T_{b}$ and $H_{a}$ under usual assumption $\log(t_{exp}f_{0} )=20\div 25$, we need this diameter to be about 24 nm. These mean diameter values are essentially more than those expected in film fabrication process. One of possible explanations of such discrepancy is the relative simplicity of used theoretical models.

The last (but not the least) question is about the nature of intergranular interaction in the studied samples. Is it of dipole-dipole or exchange nature? We do not have a convincing answer to this question. One of possible suppositions is that exchange part of the interaction coexists with a dipole-dipole one due to closeness of granules content in our film to the percolation threshold.

\section{\label{sec:concl}Conclusions}

To conclude, here we present a mean-field consideration of the magnetization of ensemble of interacting Stoner-Wohlfarth particles. We do that on the base of the kinetic equation solution. The equation has been written for the relaxation of overall ensemble magnetization to its self-consistent equilibrium state in the effective field consisting of external and the interaction fields. The latter field, in turn, is proportional to instantaneous value of overall magnetization. Numerical solution of the above kinetic equation shows that the presence of mean-field interparticle interaction leads to the following effects:

- At certain temperature, $T_{sf}$, proportional to interaction parameter $\lambda$, the system of FM granules undergoes the intergranular magnetic ordering - "superferromagnetism", yielding the additional coercivity at $T<T_{sf}$. For $T_{sf}>T_{b}$, the essential coercivity arises at temperatures above blocking temperature;

-  At $T<T_{sf}$ a coercive field increases and magnetization reversal becomes "harder", i.e. the hysteresis loops become almost rectangular with increased remanence;

-  Temperature dependence of coercive field in the low temperature region resembles very much Neel-Broun law; for $T_{sf}>T_{b}$ at increasing temperature the dependence $H_{c}(\sqrt{T})$ (or $h_{c}(\sqrt{T_{red}})$ in the dimensionless units) has an inflexion point at blocking temperature and then continues to go linearly in $\sqrt{T}$ up to $T_{sf}$ with much smaller slope;

- In the system of weakly interacting particles the values of $T_{b}$ and $H_{c}(T)$ at $T<T_{b}$ depend on the measuring time as it is usual for SW particles. At the same time, in the case of $T_{sf}>T_{b}$ the dependence $H_{c}(\sqrt{T})$ in the range $T_{b}<T<T_{sf}$ ceases to depend on measuring time.

- The temperature dependence of a coercive field (Eq. (\ref{EQ8})) is described with good accuracy by two additive contributions. The first one (below blocking temperature) is strongly dependent on measuring time and reflects coercivity related to a metastability of the system at finite measuring times. The second one, which is almost independent from measuring time, has a temperature dependence described by Eq. (\ref{EQ7}). This contribution reflects a change in a mean field of intergranular interaction in the process of magnetization reversal.

All above manifestations of interparticle interaction in SW particles ensemble are observed experimentally in magnetostatic (with 64 times difference in measuring times) measurements of a magnetic field and temperature dependencies of magnetization of (Co$_{0.25}$Fe$_{0.66}$B$_{0.09}$)$_{0.55} - $(SiO$_{2}$)$_{0.45}$ nanogranular films with FM granules content close, but below a percolation threshold. Thus it is firmly established that within the described approach the results of our numerical modeling, are in good coincidence with the experimental data.

\begin{acknowledgments}
 This work was partly supported by the grant of NAS of Ukraine Target Program "Nanostructural systems, nanomaterials and nanotechnologies".
\end{acknowledgments}

\bibliography{my}

\begin{thebibliography}{27}
\expandafter\ifx\csname natexlab\endcsname\relax\def\natexlab#1{#1}\fi
\expandafter\ifx\csname bibnamefont\endcsname\relax
  \def\bibnamefont#1{#1}\fi
\expandafter\ifx\csname bibfnamefont\endcsname\relax
  \def\bibfnamefont#1{#1}\fi
\expandafter\ifx\csname citenamefont\endcsname\relax
  \def\citenamefont#1{#1}\fi
\expandafter\ifx\csname url\endcsname\relax
  \def\url#1{\texttt{#1}}\fi
\expandafter\ifx\csname urlprefix\endcsname\relax\def\urlprefix{URL }\fi
\providecommand{\bibinfo}[2]{#2}
\providecommand{\eprint}[2][]{\url{#2}}

\bibitem[{\citenamefont{Mao et~al.}(2008)\citenamefont{Mao, Chen, and He}}]{r1}
\bibinfo{author}{\bibfnamefont{Z.}~\bibnamefont{Mao}},
  \bibinfo{author}{\bibfnamefont{D.}~\bibnamefont{Chen}}, \bibnamefont{and}
  \bibinfo{author}{\bibfnamefont{Z.}~\bibnamefont{He}}, \bibinfo{journal}{J.
  Magn. Magn. Mater.} \textbf{\bibinfo{volume}{320}}, \bibinfo{pages}{642}
  (\bibinfo{year}{2008}).

\bibitem[{\citenamefont{Liu and Bertram}(2001)}]{r2}
\bibinfo{author}{\bibfnamefont{A.~D.} \bibnamefont{Liu}} \bibnamefont{and}
  \bibinfo{author}{\bibfnamefont{H.~N.} \bibnamefont{Bertram}},
  \bibinfo{journal}{J. Appl. Phys.} \textbf{\bibinfo{volume}{89}},
  \bibinfo{pages}{2861} (\bibinfo{year}{2001}).

\bibitem[{\citenamefont{Binns et~al.}(2002)\citenamefont{Binns, Maher,
  Kechrakos, and Trohidou}}]{r3}
\bibinfo{author}{\bibfnamefont{C.}~\bibnamefont{Binns}},
  \bibinfo{author}{\bibfnamefont{M.~J.} \bibnamefont{Maher}},
  \bibinfo{author}{\bibfnamefont{D.}~\bibnamefont{Kechrakos}},
  \bibnamefont{and} \bibinfo{author}{\bibfnamefont{K.~N.}
  \bibnamefont{Trohidou}}, \bibinfo{journal}{Phys. Rev. B}
  \textbf{\bibinfo{volume}{66}}, \bibinfo{pages}{184413}
  (\bibinfo{year}{2002}).

\bibitem[{\citenamefont{Allia et~al.}(2001)\citenamefont{Allia, Coisson,
  Tiberto, Vinai, Knobel, Novak, and Nunes}}]{r4}
\bibinfo{author}{\bibfnamefont{P.}~\bibnamefont{Allia}},
  \bibinfo{author}{\bibfnamefont{M.}~\bibnamefont{Coisson}},
  \bibinfo{author}{\bibfnamefont{P.}~\bibnamefont{Tiberto}},
  \bibinfo{author}{\bibfnamefont{F.}~\bibnamefont{Vinai}},
  \bibinfo{author}{\bibfnamefont{M.}~\bibnamefont{Knobel}},
  \bibinfo{author}{\bibfnamefont{M.~A.} \bibnamefont{Novak}}, \bibnamefont{and}
  \bibinfo{author}{\bibfnamefont{W.~C.} \bibnamefont{Nunes}},
  \bibinfo{journal}{Phys. Rev. B} \textbf{\bibinfo{volume}{64}},
  \bibinfo{pages}{144420} (\bibinfo{year}{2001}).

\bibitem[{\citenamefont{Escrig et~al.}(2008)\citenamefont{Escrig, Allende,
  Altbir, and Bahiana}}]{r5}
\bibinfo{author}{\bibfnamefont{J.}~\bibnamefont{Escrig}},
  \bibinfo{author}{\bibfnamefont{S.}~\bibnamefont{Allende}},
  \bibinfo{author}{\bibfnamefont{D.}~\bibnamefont{Altbir}}, \bibnamefont{and}
  \bibinfo{author}{\bibfnamefont{M.}~\bibnamefont{Bahiana}},
  \bibinfo{journal}{Appl. Phys. Lett.} \textbf{\bibinfo{volume}{93}},
  \bibinfo{pages}{023101} (\bibinfo{year}{2008}).

\bibitem[{\citenamefont{Hillenkamp et~al.}(2008)\citenamefont{Hillenkamp,
  Domenicantonio, and Felix}}]{r6}
\bibinfo{author}{\bibfnamefont{M.}~\bibnamefont{Hillenkamp}},
  \bibinfo{author}{\bibfnamefont{G.}~\bibnamefont{Domenicantonio}},
  \bibnamefont{and} \bibinfo{author}{\bibfnamefont{C.}~\bibnamefont{Felix}},
  \bibinfo{journal}{Phys. Rev. B} \textbf{\bibinfo{volume}{77}},
  \bibinfo{pages}{014422} (\bibinfo{year}{2008}).

\bibitem[{\citenamefont{Yao et~al.}(2008)\citenamefont{Yao, Ge, Zhou, and
  Zuo}}]{r7}
\bibinfo{author}{\bibfnamefont{D.}~\bibnamefont{Yao}},
  \bibinfo{author}{\bibfnamefont{S.}~\bibnamefont{Ge}},
  \bibinfo{author}{\bibfnamefont{X.}~\bibnamefont{Zhou}}, \bibnamefont{and}
  \bibinfo{author}{\bibfnamefont{H.}~\bibnamefont{Zuo}}, \bibinfo{journal}{J.
  Appl. Phys.} \textbf{\bibinfo{volume}{104}}, \bibinfo{pages}{013902}
  (\bibinfo{year}{2008}).

\bibitem[{\citenamefont{Luttinger and Tisza}(1946)}]{r25}
\bibinfo{author}{\bibfnamefont{J.~M.} \bibnamefont{Luttinger}}
  \bibnamefont{and} \bibinfo{author}{\bibfnamefont{L.}~\bibnamefont{Tisza}},
  \bibinfo{journal}{Phys. Rev.} \textbf{\bibinfo{volume}{70}},
  \bibinfo{pages}{954} (\bibinfo{year}{1946}).

\bibitem[{\citenamefont{Meilikhov and Farzetdinova}(2002)}]{r26}
\bibinfo{author}{\bibfnamefont{E.~Z.} \bibnamefont{Meilikhov}}
  \bibnamefont{and} \bibinfo{author}{\bibfnamefont{R.~M.}
  \bibnamefont{Farzetdinova}}, \bibinfo{journal}{JETP}
  \textbf{\bibinfo{volume}{94}}, \bibinfo{pages}{751} (\bibinfo{year}{2002}).

\bibitem[{\citenamefont{Pogorelov et~al.}(2008)\citenamefont{Pogorelov,
  Kakazei, Costa, and Sousa}}]{r24}
\bibinfo{author}{\bibfnamefont{Y.~G.} \bibnamefont{Pogorelov}},
  \bibinfo{author}{\bibfnamefont{G.~N.} \bibnamefont{Kakazei}},
  \bibinfo{author}{\bibfnamefont{M.~D.} \bibnamefont{Costa}}, \bibnamefont{and}
  \bibinfo{author}{\bibfnamefont{J.~B.} \bibnamefont{Sousa}},
  \bibinfo{journal}{J. Appl. Phys.} \textbf{\bibinfo{volume}{103}},
  \bibinfo{pages}{07B723} (\bibinfo{year}{2008}).

\bibitem[{\citenamefont{S.V.Vonsovsky}(1974)}]{r8}
\bibinfo{author}{\bibnamefont{S.V.Vonsovsky}},
  \emph{\bibinfo{title}{Magnetizm}} (\bibinfo{publisher}{John Wiley, New York},
  \bibinfo{year}{1974}).

\bibitem[{\citenamefont{Iskhakov et~al.}(2004)\citenamefont{Iskhakov, Frolov,
  Zhigalov, and Procof`ev}}]{r9}
\bibinfo{author}{\bibfnamefont{R.~S.} \bibnamefont{Iskhakov}},
  \bibinfo{author}{\bibfnamefont{G.~I.} \bibnamefont{Frolov}},
  \bibinfo{author}{\bibfnamefont{V.~S.} \bibnamefont{Zhigalov}},
  \bibnamefont{and} \bibinfo{author}{\bibfnamefont{D.~J.}
  \bibnamefont{Procof`ev}}, \bibinfo{journal}{Lett. J. Tech. Phys.}
  \textbf{\bibinfo{volume}{30}}, \bibinfo{pages}{51} (\bibinfo{year}{2004}).

\bibitem[{\citenamefont{Jonsson et~al.}(1996)\citenamefont{Jonsson, Turkki,
  Strom, El-Shall, and Rao}}]{r10}
\bibinfo{author}{\bibfnamefont{B.~J.} \bibnamefont{Jonsson}},
  \bibinfo{author}{\bibfnamefont{T.}~\bibnamefont{Turkki}},
  \bibinfo{author}{\bibfnamefont{V.}~\bibnamefont{Strom}},
  \bibinfo{author}{\bibfnamefont{M.~S.} \bibnamefont{El-Shall}},
  \bibnamefont{and} \bibinfo{author}{\bibfnamefont{K.~V.} \bibnamefont{Rao}},
  \bibinfo{journal}{J. Appl. Phys.} \textbf{\bibinfo{volume}{79}},
  \bibinfo{pages}{5063} (\bibinfo{year}{1996}).

\bibitem[{\citenamefont{Perez et~al.}(1995)\citenamefont{Perez, Dupuis,
  Tuaillon, Perez, Paillard, Melinon, Treilleux, Barbara, Thomas, and
  Boushet-Fabrer}}]{r11}
\bibinfo{author}{\bibfnamefont{J.~P.} \bibnamefont{Perez}},
  \bibinfo{author}{\bibfnamefont{V.}~\bibnamefont{Dupuis}},
  \bibinfo{author}{\bibfnamefont{J.}~\bibnamefont{Tuaillon}},
  \bibinfo{author}{\bibfnamefont{A.}~\bibnamefont{Perez}},
  \bibinfo{author}{\bibfnamefont{V.}~\bibnamefont{Paillard}},
  \bibinfo{author}{\bibfnamefont{P.}~\bibnamefont{Melinon}},
  \bibinfo{author}{\bibfnamefont{M.}~\bibnamefont{Treilleux}},
  \bibinfo{author}{\bibfnamefont{B.}~\bibnamefont{Barbara}},
  \bibinfo{author}{\bibfnamefont{L.}~\bibnamefont{Thomas}}, \bibnamefont{and}
  \bibinfo{author}{\bibfnamefont{B.}~\bibnamefont{Boushet-Fabrer}},
  \bibinfo{journal}{J. Magn. Magn. Mater.} \textbf{\bibinfo{volume}{145}},
  \bibinfo{pages}{74} (\bibinfo{year}{1995}).

\bibitem[{\citenamefont{Kleemann et~al.}(2001)\citenamefont{Kleemann, Petracic,
  Binek, Kakazei, Pogorelov, Sousa, Cardoso, and Freitas}}]{r12}
\bibinfo{author}{\bibfnamefont{W.}~\bibnamefont{Kleemann}},
  \bibinfo{author}{\bibfnamefont{O.}~\bibnamefont{Petracic}},
  \bibinfo{author}{\bibfnamefont{C.}~\bibnamefont{Binek}},
  \bibinfo{author}{\bibfnamefont{G.~N.} \bibnamefont{Kakazei}},
  \bibinfo{author}{\bibfnamefont{Y.~G.} \bibnamefont{Pogorelov}},
  \bibinfo{author}{\bibfnamefont{J.~B.} \bibnamefont{Sousa}},
  \bibinfo{author}{\bibfnamefont{S.}~\bibnamefont{Cardoso}}, \bibnamefont{and}
  \bibinfo{author}{\bibfnamefont{P.~P.} \bibnamefont{Freitas}},
  \bibinfo{journal}{Phys. Rev. B} \textbf{\bibinfo{volume}{63}},
  \bibinfo{pages}{134423} (\bibinfo{year}{2001}).

\bibitem[{\citenamefont{Chen et~al.}(2004)\citenamefont{Chen, Sahoo, Kleemann,
  Cardoso, and Freitas}}]{r13}
\bibinfo{author}{\bibfnamefont{X.}~\bibnamefont{Chen}},
  \bibinfo{author}{\bibfnamefont{S.}~\bibnamefont{Sahoo}},
  \bibinfo{author}{\bibfnamefont{W.}~\bibnamefont{Kleemann}},
  \bibinfo{author}{\bibfnamefont{S.}~\bibnamefont{Cardoso}}, \bibnamefont{and}
  \bibinfo{author}{\bibfnamefont{P.~P.} \bibnamefont{Freitas}},
  \bibinfo{journal}{Phys. Rev. B} \textbf{\bibinfo{volume}{70}},
  \bibinfo{pages}{172411} (\bibinfo{year}{2004}).

\bibitem[{\citenamefont{Atherton}(1990)}]{r14}
\bibinfo{author}{\bibfnamefont{D.~L.} \bibnamefont{Atherton}},
  \bibinfo{journal}{IEEE Trans. Magn.} \textbf{\bibinfo{volume}{26}},
  \bibinfo{pages}{3059} (\bibinfo{year}{1990}).

\bibitem[{\citenamefont{Chuev and Hesse}(2007)}]{r15}
\bibinfo{author}{\bibfnamefont{M.~A.} \bibnamefont{Chuev}} \bibnamefont{and}
  \bibinfo{author}{\bibfnamefont{J.}~\bibnamefont{Hesse}}, \bibinfo{journal}{J.
  Phys.: Cond. Matt.} \textbf{\bibinfo{volume}{19}}, \bibinfo{pages}{506201}
  (\bibinfo{year}{2007}).

\bibitem[{\citenamefont{Zhong et~al.}(2005)\citenamefont{Zhong, Zhu, Guo, and
  Lin}}]{r16}
\bibinfo{author}{\bibfnamefont{J.~J.} \bibnamefont{Zhong}},
  \bibinfo{author}{\bibfnamefont{J.~G.} \bibnamefont{Zhu}},
  \bibinfo{author}{\bibfnamefont{Y.~G.} \bibnamefont{Guo}}, \bibnamefont{and}
  \bibinfo{author}{\bibfnamefont{Z.~W.} \bibnamefont{Lin}},
  \bibinfo{journal}{IEEE Trans. Magn.} \textbf{\bibinfo{volume}{41}},
  \bibinfo{pages}{1496} (\bibinfo{year}{2005}).

\bibitem[{\citenamefont{Neel}(1949)}]{r17}
\bibinfo{author}{\bibfnamefont{L.}~\bibnamefont{Neel}}, \bibinfo{journal}{Ann.
  Geophys.} \textbf{\bibinfo{volume}{5}}, \bibinfo{pages}{99}
  (\bibinfo{year}{1949}).

\bibitem[{\citenamefont{Chuev}(2007)}]{r18}
\bibinfo{author}{\bibfnamefont{M.~A.} \bibnamefont{Chuev}},
  \bibinfo{journal}{JETP Lett.} \textbf{\bibinfo{volume}{85}},
  \bibinfo{pages}{744} (\bibinfo{year}{2007}).

\bibitem[{\citenamefont{Timofeev et~al.}(2008)\citenamefont{Timofeev, Kalita,
  and Ryabchenko}}]{r19}
\bibinfo{author}{\bibfnamefont{A.~A.} \bibnamefont{Timofeev}},
  \bibinfo{author}{\bibfnamefont{V.~M.} \bibnamefont{Kalita}},
  \bibnamefont{and} \bibinfo{author}{\bibfnamefont{S.~M.}
  \bibnamefont{Ryabchenko}}, \bibinfo{journal}{Low Temp. Phys.}
  \textbf{\bibinfo{volume}{34}}, \bibinfo{pages}{446} (\bibinfo{year}{2008}).

\bibitem[{\citenamefont{Wang et~al.}(2001)\citenamefont{Wang, Ding, Kong, Li,
  and Feng}}]{r20}
\bibinfo{author}{\bibfnamefont{L.}~\bibnamefont{Wang}},
  \bibinfo{author}{\bibfnamefont{J.}~\bibnamefont{Ding}},
  \bibinfo{author}{\bibfnamefont{H.~Z.} \bibnamefont{Kong}},
  \bibinfo{author}{\bibfnamefont{Y.}~\bibnamefont{Li}}, \bibnamefont{and}
  \bibinfo{author}{\bibfnamefont{Y.~P.} \bibnamefont{Feng}},
  \bibinfo{journal}{Phys. Rev. B} \textbf{\bibinfo{volume}{64}},
  \bibinfo{pages}{214410} (\bibinfo{year}{2001}).

\bibitem[{\citenamefont{Munakata et~al.}(1999)\citenamefont{Munakata, Yagi, and
  Shimada}}]{r21}
\bibinfo{author}{\bibfnamefont{M.}~\bibnamefont{Munakata}},
  \bibinfo{author}{\bibfnamefont{M.}~\bibnamefont{Yagi}}, \bibnamefont{and}
  \bibinfo{author}{\bibfnamefont{Y.}~\bibnamefont{Shimada}},
  \bibinfo{journal}{IEEE Trans. Magn.} \textbf{\bibinfo{volume}{35}},
  \bibinfo{pages}{3430} (\bibinfo{year}{1999}).

\bibitem[{\citenamefont{Johnsson et~al.}(2003)\citenamefont{Johnsson, Aoqui,
  Grishin, and Munakata}}]{r22}
\bibinfo{author}{\bibfnamefont{P.}~\bibnamefont{Johnsson}},
  \bibinfo{author}{\bibfnamefont{S.~I.} \bibnamefont{Aoqui}},
  \bibinfo{author}{\bibfnamefont{A.~M.} \bibnamefont{Grishin}},
  \bibnamefont{and} \bibinfo{author}{\bibfnamefont{M.}~\bibnamefont{Munakata}},
  \bibinfo{journal}{J. Appl. Phys.} \textbf{\bibinfo{volume}{93}},
  \bibinfo{pages}{8101} (\bibinfo{year}{2003}).

\bibitem[{\citenamefont{Timopheev and Ryabchenko}(2008)}]{r23}
\bibinfo{author}{\bibfnamefont{A.~A.} \bibnamefont{Timopheev}}
  \bibnamefont{and} \bibinfo{author}{\bibfnamefont{S.~M.}
  \bibnamefont{Ryabchenko}}, \bibinfo{journal}{Ukr. J. Phys.}
  \textbf{\bibinfo{volume}{53}}, \bibinfo{pages}{261} (\bibinfo{year}{2008}).

\bibitem[{\citenamefont{Munakata et~al.}(2002)\citenamefont{Munakata, Namikawa,
  Motoyama, Yagi, Shimada, Yamaguchi, and Arai}}]{r27}
\bibinfo{author}{\bibfnamefont{M.}~\bibnamefont{Munakata}},
  \bibinfo{author}{\bibfnamefont{M.}~\bibnamefont{Namikawa}},
  \bibinfo{author}{\bibfnamefont{M.}~\bibnamefont{Motoyama}},
  \bibinfo{author}{\bibfnamefont{M.}~\bibnamefont{Yagi}},
  \bibinfo{author}{\bibfnamefont{Y.}~\bibnamefont{Shimada}},
  \bibinfo{author}{\bibfnamefont{M.}~\bibnamefont{Yamaguchi}},
  \bibnamefont{and} \bibinfo{author}{\bibfnamefont{K.}~\bibnamefont{Arai}},
  \bibinfo{journal}{J. Magn. Soc. Japan} \textbf{\bibinfo{volume}{26}},
  \bibinfo{pages}{388} (\bibinfo{year}{2002}).

\end{thebibliography}

\end{document}